\documentclass[lettersize,journal]{IEEEtran}
\usepackage{amsmath,amsfonts}
\usepackage{algorithmic}
\usepackage{algorithm}
\usepackage{array}
\usepackage[caption=false,font=normalsize,labelfont=sf,textfont=sf]{subfig}
\usepackage{textcomp}
\usepackage{stfloats}
\usepackage{float}
\usepackage{url}
\usepackage{verbatim}
\usepackage{graphicx}
\usepackage{cite}
\usepackage{siunitx}
\usepackage{balance}
\usepackage{circuitikz}
\usepackage{capt-of}
\usepackage[export]{adjustbox}
\usepackage{subfig}
\newcommand{\etal}{\textit{et al}}
\newcommand{\ignore}[1]{}

\begin{document}
\bstctlcite{IEEEexample:BSTcontrol}

	\title{An Edge-Coupled Magnetostatic Bandpass Filter}
	
	\author{Connor Devitt,~\IEEEmembership{Graduate Student Member,~IEEE}, Renyuan Wang,~\IEEEmembership{Member,~IEEE},\\Sudhanshu Tiwari,~\IEEEmembership{Member,~IEEE}, Sunil A. Bhave,~\IEEEmembership{Senior Member,~IEEE}
\thanks{Manuscript received on XX XX, 2023; revised on XX XX, 2024; accepted on XX XX, 2024. This research was developed with funding from the Air Force Research Laboratory (AFRL) and the Defense Advanced Research Projects Agency (DARPA). The views, opinions and/or findings expressed are those of the authors and should not be interpreted as representing the official views or policies of the Department of Defense or the U.S. Government. This manuscript is approved for public release; distribution A: distribution unlimited. (\textit{Corresponding authors: Connor Devitt, Renyuan Wang})}	

\thanks{R.W. invented the device concept, and developed baseline model and device design. C.D. performed simulations on fabrication variations and total filter loss, chip fabrication and characterization, as well as data analysis. Manuscript was prepared by C.D. with inputs from R.W., S.T., and S.A.B.}

  \thanks{Connor Devitt (e-mail: devitt@purdue.edu), Sudhanshu Tiwari (e-mail: tiwari40@purdue.edu), and Sunil A. Bhave (e-mail: bhave@purdue.edu) are with the OxideMEMS Lab, Elmore Family School of Electrical and Computer Engineering, Purdue University, West Lafayette, IN 47907 USA.}
		\thanks{Renyuan Wang is with FAST Labs, BAE Systems, Inc., Nashua,
			NH 03060 USA (e-mail: renyuan.wang@baesystems.com).}}
	
	{}
	
	\maketitle
	
	\begin{abstract}
		This paper reports on the design, fabrication, and characterization of an edge-coupled magnetostatic forward volume wave bandpass filter. Using micromachining techniques, the filter is fabricated from a yttrium iron garnet (YIG) film grown on a gadolinium gallium garnet (GGG) substrate with inductive transducers. By adjusting an out-of-plane magnetic field, we demonstrate linear center frequency tuning for a 4\textsuperscript{th}-order filter from 4.5 GHz to 10.1 GHz while retaining a fractional bandwidth of 0.3\%, an insertion loss of 6.94 dB, and a -35dB rejection. We characterize the filter nonlinearity in the passband and stopband with IIP3 measurements of -4.85 dBm and 25.84 dBm, respectively. When integrated with a tunable magnetic field, this device is an octave tunable narrowband channel-select filter.
	\end{abstract}
	
	
	\section{Introduction}
	\IEEEPARstart{C}{oupled} micro-electromechanical resonators with high quality factors (Q-factor) and miniaturized footprints have been an attractive technology for integration in wireless communication systems as narrowband channel-select filters. Wang \etal \cite{MEMS:wang} has demonstrated high-order micromechanical bandpass filters using one-dimensional (1D) arrays of mechanically-coupled resonators. However, for higher-order filters, long mechanical coupling beams become impractical due to size constraints \cite{MEMS:edge_couple} while sensitivity due to fabrication variation causes increased insertion loss and passband distortion \cite{MEMS:weinstein}. The high sensitivity of weak electrostatic or mechanical edge-coupled resonators due to structural asymmetries has been leveraged for sensitive parametric mass sensing applications in \cite{MEMS:edge_couple}, but prohibits their use in filters. 2D microresonator arrays have shown some success by utilizing weak coupling in one dimension to achieve pass band shape and strong coupling in the other dimension to reduce effects of fabrication variation, but suffer from high insertion loss \cite{MEMS:weinstein}.
    Coupled resonator arrays utilizing magnetostatic waves (MSW) have the potential to overcome the weak coupling and extremely narrow bandwidths achieved by electrostatically coupled micromechanical resonators \cite{pourkamali_electrostatically_2004,alastalo_systematic_2006} while introducing a degree of tunability.
    
	\begin{figure}[t]
		\vspace{-0.1in}
		\begin{center} 
			\noindent
			\subfloat[]{\includegraphics[width=2.8in]{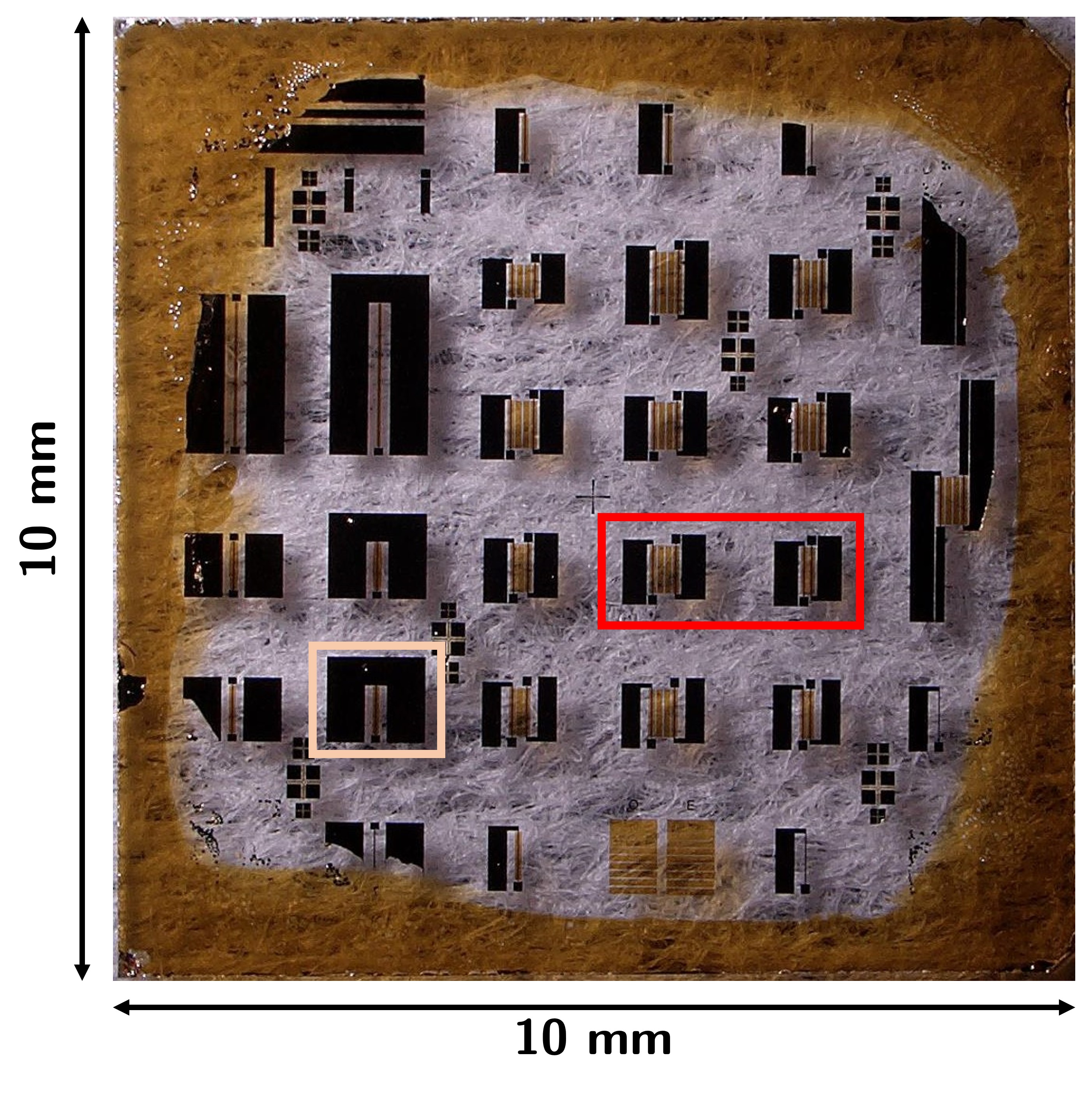}%
				\label{whole_chip}}
			\hfil
			\subfloat[]{\includegraphics[width=2.8in]{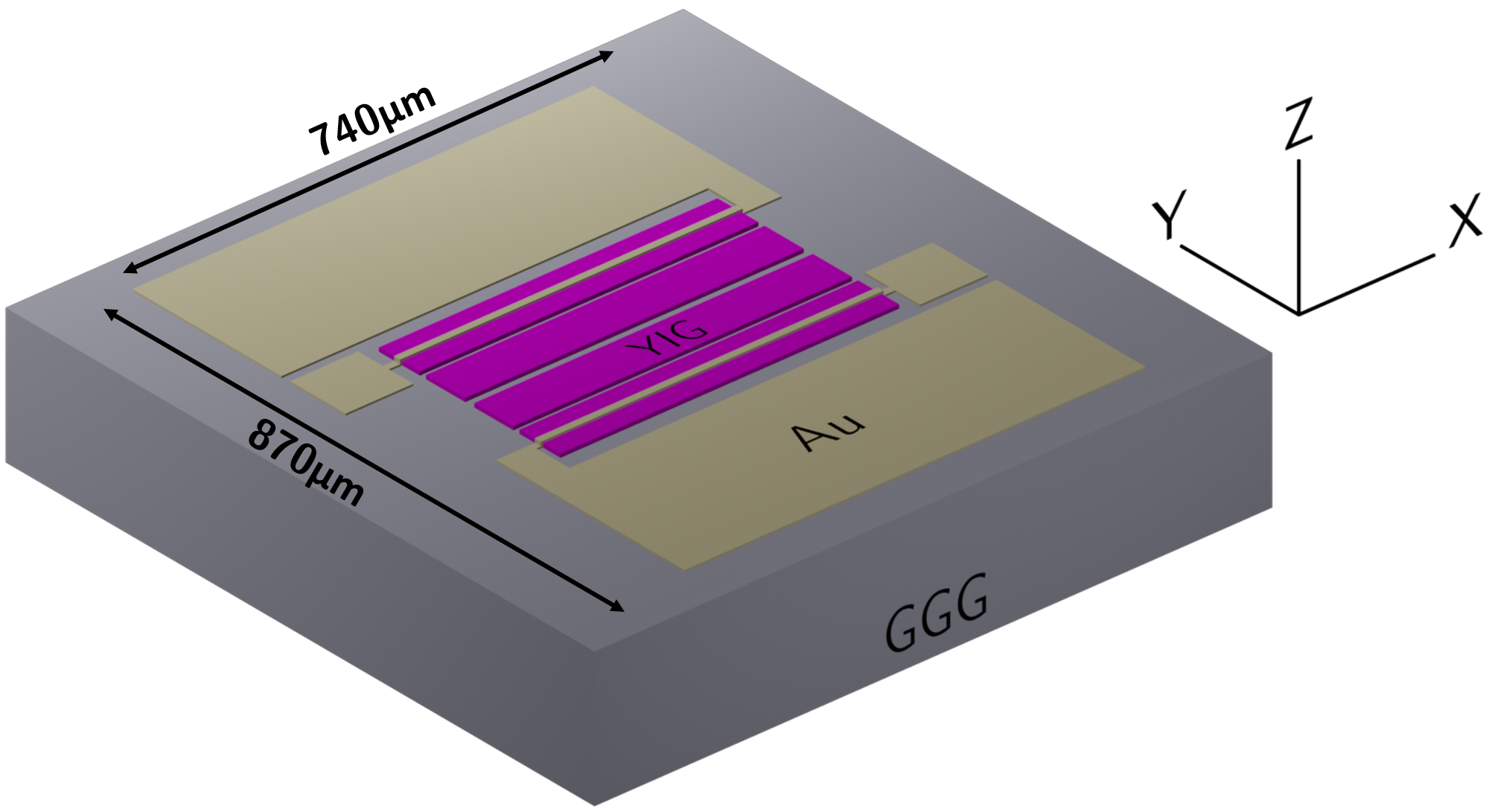}%
				\label{chip_cartoon}}
			
			\caption{\textbf{(a)} Chip microphotograph of multiple MSW bandpass filters and 1-port resonators fabricated on a YIG on GGG chip using YIG micromachining technology \cite{MSFVW:yiyang}. A 4-pole filter (on the left) and a 2-pole filter (on the right) are highlighted in red. A 1-port resonator is highlighted in orange. \textbf{(b)} Rendering of a 4-pole bandpass filter featuring four YIG resonators with gold electrodes conformally deposited over the etched YIG. Magnetic bias is oriented out-of-plane along the $z$-axis. }
			\label{chip_pictures}
		\end{center}
		\vspace{-0.15in}
	\end{figure}

     The magnetostatic wave resonance can be tuned more than an octave in frequency using a static magnetic field ensuring that the filter size does not scale to sub-micrometer dimensions at high-frequencies. Yttrium iron garnet (YIG) is the most widely used material for MSW devices due to its low Gilbert damping ($\alpha = 2.8\times10^{-4}$ for a $\SI{100}{\nano\meter}$ film \cite{MSFVW:YOGdampling}) and experimentally demonstrated Q-factors exceeding $3000$ \cite{MSFVW:Dai, MSW:marcelli1991tunable}. In state of the art YIG sphere filters \cite{micro_lambda_wireless_inc_mlfd_nodate, teledyne_yig_sphere, micro_lambda_wireless_inc_yig_spheres}, polished YIG resonators are attached to a thermally conductive rod and manually aligned to non-planar inductive loops acting as transducers. The assembled sphere and loop structures are then coupled through transmission lines similar to the coupling beams in \cite{MEMS:wang} to synthesize a filter. Planar YIG resonators can magnetically couple if they are fabricated in close proximity (Fig. $\ref{chip_cartoon}$), analogous to the electrically coupled mechanical resonators in \cite{MEMS:edge_couple,MEMS:masssensor} or coupled electromagnetic resonators. Magnetically coupled YIG filters can be fabricated at scale using micromachining techniques on films grown on a gadolinium gallium garnet (GGG) substrate, allowing for miniaturization and eliminating the need for polishing and meticulous manual alignment.
    	
    \section{Bandpass Filter Design}

	The bandpass filter shown in Fig. $\ref{chip_cartoon}$ consists of a number of closely-spaced rectangular YIG magnetostatic forward volume wave (MSFVW) resonators with shorted $\SI{300}{\nano\meter}$-thick gold inductive transducers conformally deposited over the outermost resonators. With an out-of-plane DC magnetic bias, the RF magnetic field from the transducers excite MSFVW modes in the YIG mesa. Forward volume waves are a family of highly dispersive modes in a thin film whose lowest order mode is described by the dispersion relation \cite{MSFVW:Stancil}:
	\begin{equation}
		\omega^2 = \omega_0\left[\omega_0+\omega_m\left(1 - \frac{1-e^{-k_{mn}t}}{k_{mn}t}\right)\right],
	\end{equation}
	where $\omega_m = \mu_0\gamma_m M_s$, $\omega_0 =\mu_0\gamma_m H^{eff}_{DC}$, $t$ is the film thickness, $\gamma_m$ is the gyromagnetic ratio, $\mu_0$ is the permeability of free space, $k_{mn}$ is the wave vector, $M_s$ is the saturation magnetization, and $H^{eff}_{DC}$ is the effective DC magnetic field. Considering the limits as $k_{mn}\rightarrow0$ and $k_{mn}\rightarrow\infty$ in the dispersion relation, $\omega$ is restricted within the range \cite{MSFVW:Stancil}:
	\begin{equation}
		\omega_0 \leq \omega \leq \sqrt{\omega_0\left(\omega_0+\omega_m\right)},
		\label{spinwave_manifold}
	\end{equation}
    denoted as the spin wave manifold. When the planar dimensions of the thin YIG film are bounded, the magnetostatic waves reflects off the edges forming a standing waves with wave vectors approximately given by \cite{MSFVW:Ishak, MSFVW:ishak1988tunable,MSFVW:Hanna,MSFVW:Marcelli2, MSFVW:marcelli1996magnetostatic}:
	\begin{equation}
		k_{mn} = \sqrt{\left(\frac{\pi m}{l}\right)^2 + \left(\frac{\pi n}{w}\right)^2}, \ m,n=1,2,3\dots
	\end{equation}
	where $l$ and $w$ are the length and width of the cavity respectively. These MSFVW resonances can be further understood through an analogy to Lamb waves in a piezoelectric plate \cite{wang_design_2015, BAW:hashimoto, esteves_al068sc032n_2021, giribaldi_620_2023}. An oscillating electric field perturbs the polarization of the piezoelectric film generating a stress field and exciting an acoustic wave. In the ferrimagnetic film, an oscillating magnetic field, conversely, perturbs the static magnetization leading to a precession of spins. Both the piezoelectric and magnetostatic cavities support a discrete number modes whose wave vectors depend on the cavity dimensions. Nonlinear dispersion relates the wave vectors to the resonant frequencies for both MSFVW and Lamb waves. For MSFVW, this leads to irregularly spaced modes which all reside in the spin wave manifold.

    Unique to MSW, the applied out-of-plane magnetic field shifts the MSFVW dispersion relation in frequency where the tuning rate for the fundamental mode is given by
	\begin{equation}
		\frac{\partial \omega}{\partial H^{eff}_{DC}} = \mu_0\gamma_m \frac{2\omega_0 + \omega_m\left(1-\frac{1-e^{-k_{mn}t}}{k_{mn}t}\right)}{2\sqrt{\omega_0\left[\omega_0 + \omega_m\left(1-\frac{1-e^{-k_{mn}t}}{k_{mn}t}\right)\right]}},\\
	\end{equation}
	which simplifies to $\mu_0\gamma_m = \SI{2.8}{\mega\hertz / Oe}$ (for YIG) when $k_{mn}t\ll1$. The magnetostatic scalar potential, $\psi$, decays exponentially outside the YIG mesa \cite{MSFVW:Stancil, zhang_nonreciprocal_2020} so the MSFVW resonance in one YIG mesa may couple to adjacent mesas if there is sufficient overlap in their scalar potentials \cite{MSW:marcelli2005micromachined}. Consequently, the spacing of adjacent resonators and their vertical sidewall profile are critical to control the inter-resonator coupling strength. For the 4-pole filter in Fig. $\ref{chip_cartoon}$, the resonator spacings are $s_1=\SI{10}{\micro\meter}$, $s_2=\SI{15}{\micro\meter}$, and $s_3=\SI{10}{\micro\meter}$. Each has length $l=\SI{500}{\micro\meter}$. The outermost resonators have a width $w_{1,4}=\SI{70}{\micro\meter}$ while the inner resonators are slightly narrower at $w_{2,3}=\SI{67}{\micro\meter}$ which was found to marginally improve insertion loss and higher order width mode suppression based on finite element simulation.
	
	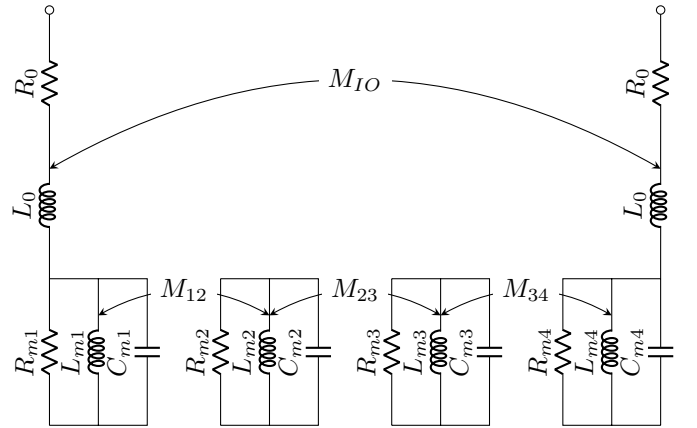
\begin{figure}[!b]
		\centering
		\begin{circuitikz}[scale=0.65]
			\ctikzset{label/align=rotate}
			
			\ctikzset{capacitors/scale=0.4}
			\ctikzset{resistors/scale=0.5}
			\ctikzset{inductors/scale=0.65}
			
			
			\draw[<->,>=stealth] (1,1.75) to [bend left] node[pos=0.5,fill=white] {$M_{12}$} ++(3.5,0);
			\draw[<->,>=stealth] (4.5,1.75) to [bend left] node[pos=0.5,fill=white] {$M_{23}$} ++(3.5,0);
			\draw[<->,>=stealth] (8,1.75) to [bend left] node[pos=0.5,fill=white] {$M_{34}$} ++(3.5,0);
			\draw[<->,>=stealth] (0,4.75) to [bend left] node[pos=0.5,fill=white] {$M_{IO}$} ++(12.5,0);
			
			\draw
			(0,3) to[L=$L_0$] (0,5)
			to[R=$R_0$, -o] (0,8)
			
			(0,0) to[R=$R_{m1}$] (0,2)
			(1,0) to[L=$L_{m1}$] (1,2)
			(2,0) to[C=$C_{m1}$] (2,2)
			
			(3.5,0) to[R=$R_{m2}$] (3.5,2)
			(4.5,0) to[L=$L_{m2}$] (4.5,2)
			(5.5,0) to[C=$C_{m2}$] (5.5,2)
			
			(7,0) to[R=$R_{m3}$] (7,2)
			(8,0) to[L=$L_{m3}$] (8,2)
			(9,0) to[C=$C_{m3}$] (9,2)
			
			(10.5,0) to[R=$R_{m4}$] (10.5,2)
			(11.5,0) to[L=$L_{m4}$] (11.5,2)
			(12.5,0) to[C=$C_{m4}$] (12.5,2)
			
			(12.5,3) to[L=$L_0$] (12.5,5)
			to[R=$R_0$, -o] (12.5,8)
			
			(0,2) -- (0,3)
			(1,2) -- (1,2.5)
			(2,2) -- (2,2.5) -- (0,2.5)
			
			(0,0) -- (0,-0.5)
			(1,0) -- (1,-0.5)
			(2,0) -- (2,-0.5) -- (0,-0.5)
			
			(12.5,2) -- (12.5,3)
			(11.5,2) -- (11.5,2.5)
			(10.5,2) -- (10.5,2.5) -- (12.5,2.5)
			
			(12.5,0) -- (12.5,-0.5)
			(11.5,0) -- (11.5,-0.5)
			(10.5,0) -- (10.5,-0.5) -- (12.5,-0.5)
			
			(3.5,2) -- (3.5,2.5)
			(4.5,2) -- (4.5,2.5)
			(5.5,2) -- (5.5,2.5) -- (3.5,2.5)
			
			(3.5,0) -- (3.5,-0.5)
			(4.5,0) -- (4.5,-0.5)
			(5.5,0) -- (5.5,-0.5) -- (3.5,-0.5)
			
			(7,2) -- (7,2.5)
			(8,2) -- (8,2.5)
			(9,2) -- (9,2.5) -- (7,2.5)
			
			(7,0) -- (7,-0.5)
			(8,0) -- (8,-0.5)
			(9,0) -- (9,-0.5) -- (7,-0.5)
			;
		\end{circuitikz}
		\caption{Lumped circuit model of an edge-coupled 4-pole MSW bandpass filter with electrically short transducers.}
		\label{lumped_circuit}
	\end{figure}

\begin{figure}[t]
        \centering
        \includegraphics[width=3.2in]{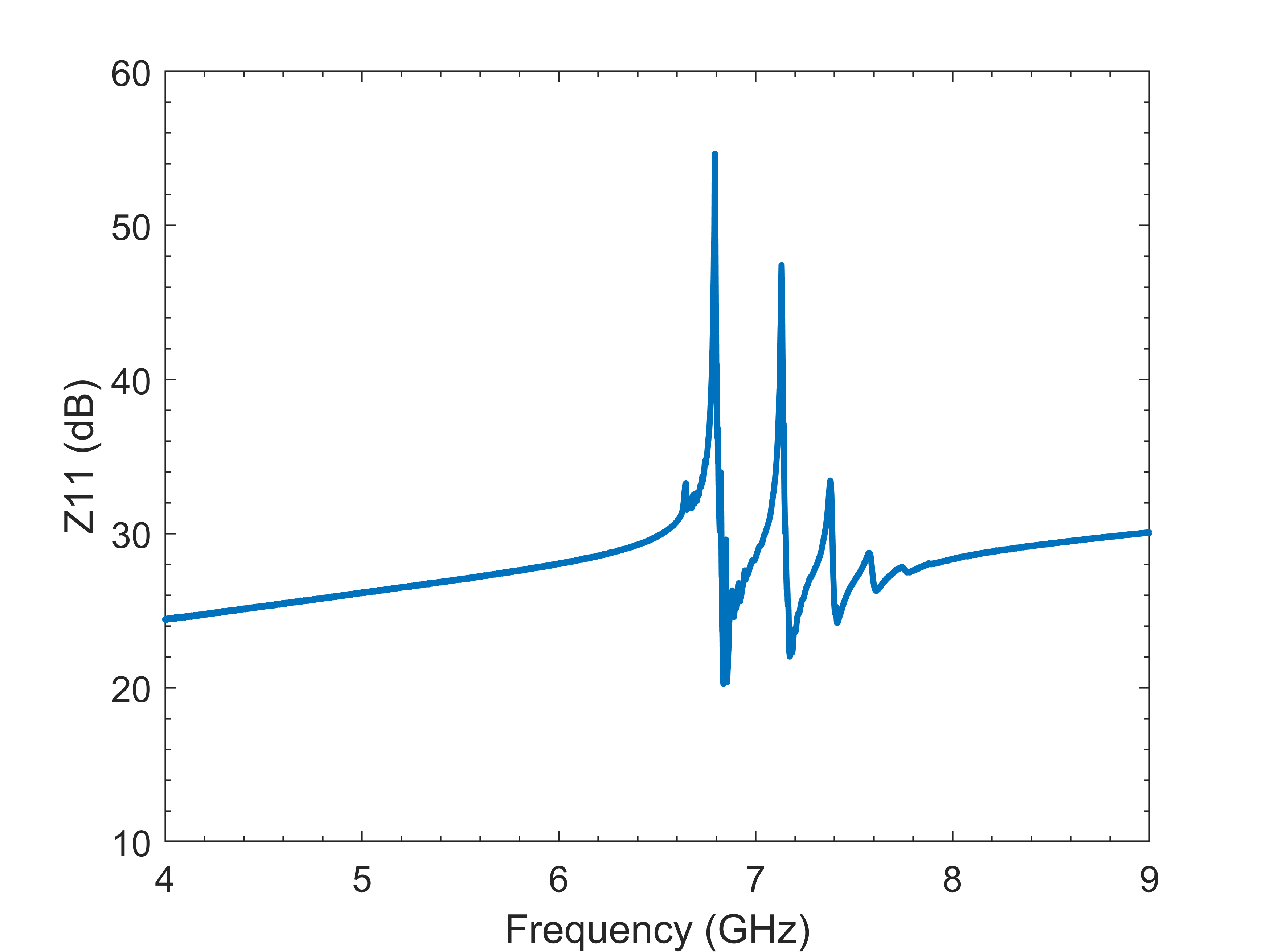}
        \caption{Measured frequency response of the 1-port resonator highlighted in Fig. $\ref{whole_chip}$ at $\SI{3962}{Oe}$ showing a $Q=2206$ and $k_{eff}^2=1.53\%$.}
        \label{resonator_response}
        \vspace*{-0.1in}
    \end{figure}

    Since each resonator length is much shorter than the electromagnetic wavelength over the tuning range, the 4-pole filter can be modeled with the lumped element circuit show in Fig. $\ref{lumped_circuit}$. A Butterworth-Van Dyke circuit \cite{FBAR:mBVD,BAW:aigner2003mems} is typically used to model an acoustic resonance where the mechanical mode is described by a series R-L-C tank circuit and the transducers introduce a shunt plate capacitance. In the case of magnetic resonators, the transducer introduces a parasitic series inductance and the MSFVW is modeled using a parallel R-L-C tank circuit instead. $L_0$ and $R_0$ represent the parasitic inductance and resistivity of the gold electrodes. $R_{m}$, $C_{m}$, and $L_{m}$ represent the MSFVW resonance of each YIG mesa. $M_{nm}$ is the inter-resonator coupling between adjacent YIG mesas while $M_{IO}$ represents input/output inductive coupling of the gold electrodes setting the out-of-band rejection level. Similar to a mechanical coupling coefficient, an effective coupling from the electrical to magnetostatic domain can be defined by ($\ref{effective_coupling}$) where $f_p$ and $f_s$ represents the magnetic resonance and anti-resonance respectively.
	\begin{equation}
		k_{eff}^2 = \frac{\pi}{2}\left(\frac{f_p}{f_s}\right)\cot\left(\frac{\pi}{2}\frac{f_p}{f_s}\right)
		\label{effective_coupling}
	\end{equation}
	$k_{eff}^2$ is a function of the resonator design determined by the ratio of $L_m$ to $L_0$. Similar to an acoustic filter, $k_{eff}^2$ sets a bound on the maximum achievable filter bandwidth and impacts the passband ripple. The representative 1-port resonator (highlighted in Fig. $\ref{whole_chip}$) exhibits a measured effective coupling and quality-factor of $k_{eff}^2=1.53\%$ and $Q=2206$ at $\SI{3962}{Oe}$ with frequency response shown in Fig. $\ref{resonator_response}$. From the measured resonator impedance response, $R_0$, $L_0$, $R_m$, $C_m$, and $L_m$ can all be extracted through separate fittings near the magnetostatic resonance and outside the spin wave manifold where the resonator behaves as an inductor. Using the same resonator parameters, the measured filter response excluding spurious modes can be fit to the lumped model in Fig. $\ref{lumped_circuit}$ by tuning $M_{IO}$, $M_{12}$, and $M_{23}$ under the assumption that each resonator has the same resonant frequency and the filter is symmetric. Fig. $\ref{4pole_fig_filter_frequency_response}$ shows the frequency response of the lumped model fitted to the measured S21 for a 4-pole filter along with the extracted model parameters.

    \section{Fabrication}
	\label{sec_fab}
	\begin{figure}[!t]
		\centering
		\includegraphics[width=3.2in]{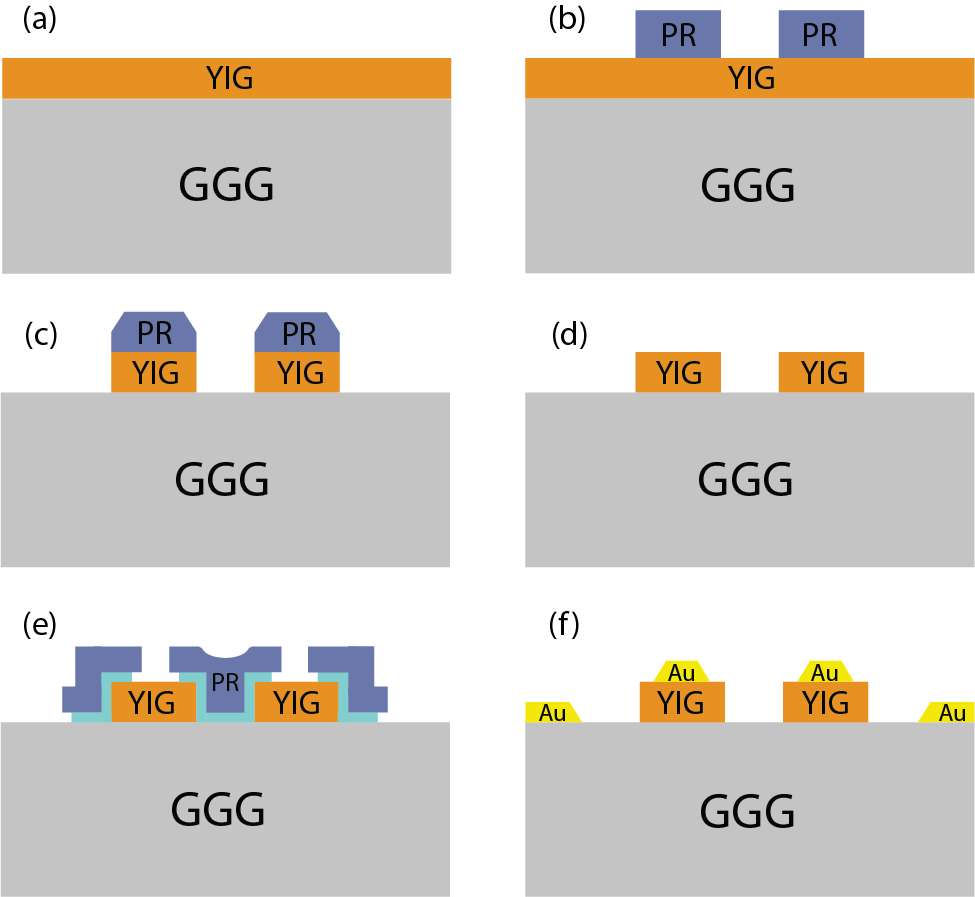}
		\caption{Fabrication process of the bandpass filter: \textbf{(a)} $\SI{3}{\micro\meter}$ liquid phase epitaxy (LPE) YIG film grown on $500\si{\micro\meter}$ GGG substrate. \textbf{(b)} $7.8\si{\micro\meter}$ thick photoresist (SPR220-7.0) patterned on YIG film as an etch mask. \textbf{(c)} $3\si{\micro\meter}$ ion mill etch of YIG film at a rate of $\SI{36}{\nano\meter / \minute}$. \textbf{(d)} Photoresist mask is removed and etched YIG is soaked in phosphoric acid at $80 \, ^\circ C$ for $ \SI{20}{\min}$. \textbf{(e)} Bi-layer photoresist (SPR220-7.0 and LOR 3B) mask for a liftoff is patterned onto the etched sample. \textbf{(f)} $\SI{10}{\nano\meter}$ Ti and $\SI{300}{\nano\meter}$ Au is deposited using glancing angle e-beam evaporation followed by a liftoff process.}
		\label{fig_fab_process}
	\end{figure}
	\begin{figure}[!t]
		\centering
		\includegraphics[width=3.2in]{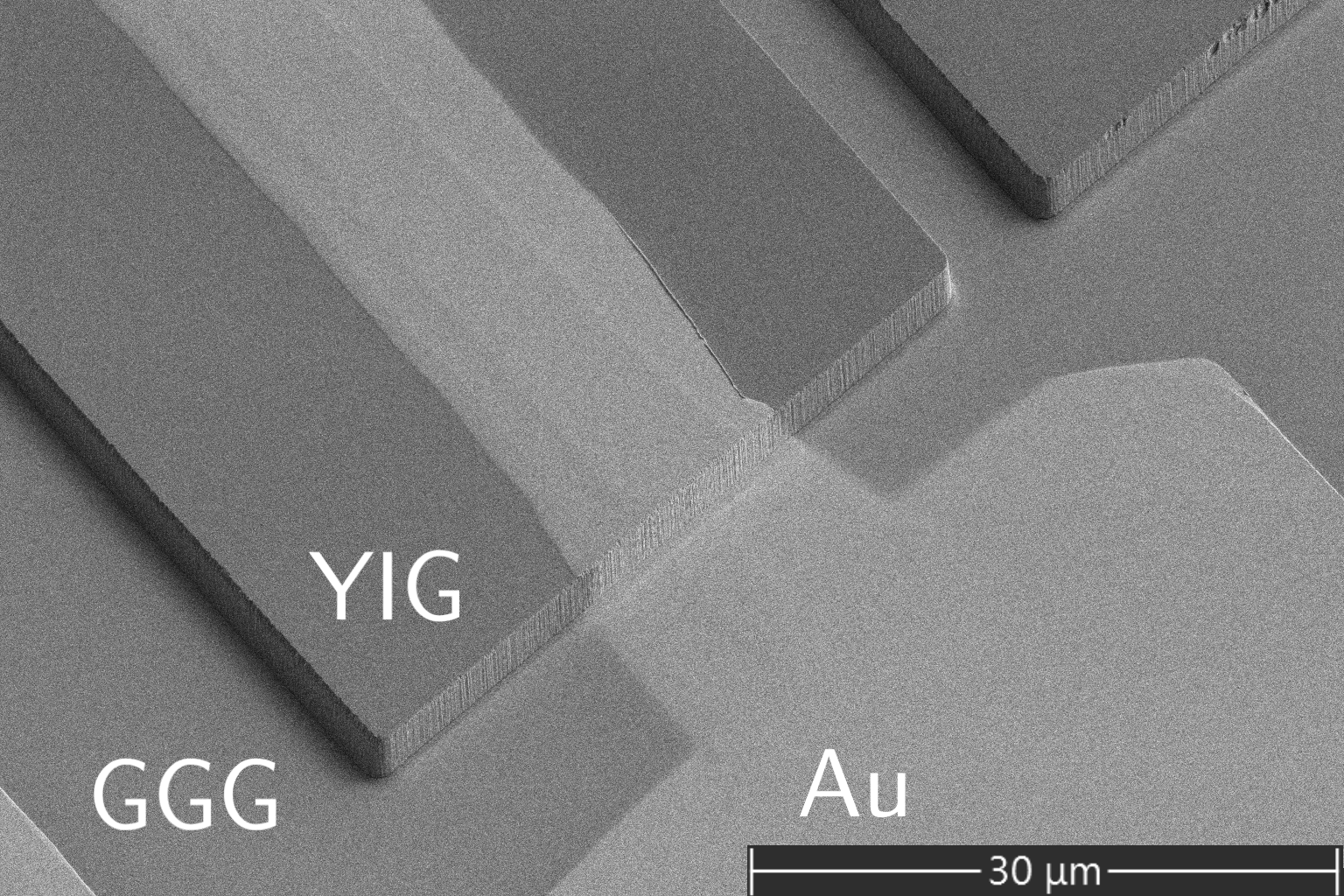}
		\caption{SEM of the MSFVW bandpass filter showing vertically etched YIG resonators with an inter-resonator spacing of \SI{10.5}{\micro\meter} and conformal gold electrodes over the edges of etched YIG.}
		\label{fig_fab_sem}
		\vspace{-0.1in}
	\end{figure}
	
	The fabrication process for the MSW bandpass filter is outlined in Fig $\ref{fig_fab_process}$ where steps (a)-(c) are adapted from \cite{MSFVW:yiyang}. A thick photoresist mask (SPR220-7.0) is patterned onto a $\SI{3}{\micro\meter}$ YIG film grown via liquid phase epitaxially (LPE) on a $\SI{500}{\micro\meter}$ GGG substrate. The YIG film is etched through at a rate of $\SI{36}{\nano\meter / \minute}$ using an optimized ion milling recipe for vertical sidewalls with sufficient intermittent cooling to prevent burning of the photoresist. Optimized lithography for the thick SPR photoresist is crucial to the filter's final performance since the inter-resonator coupling factors, $M_{ij}$, are sensitive to the physical separation of the etched YIG. With this process, we are able to fully etch resonator spacings as narrow as $\SI{3}{\micro\meter}$ setting the maximum achievable $M_{ij}\leq 1.37\%$ based on finite element simulation. After etching, the resist is removed and the sample is soaked in phosphoric acid at $80 \, ^\circ C$ to remove redeposited material. For the gold electrodes, a SPR220-7.0 photoresist mask with a liftoff resist (LOR 3B) bi-layer is patterned for a liftoff process. Using a glancing angle e-beam evaporation, $\SI{300}{\nano\meter}$ of gold and a $\SI{10}{\nano\meter}$ titanium adhesion layer is conformally deposited over the etched YIG resonators. Finally, the sample is soaked in \mbox{Remover PG} overnight to complete the liftoff process. Fig. $\ref{fig_fab_sem}$ shows an SEM of a fabricated filter illustrating the vertically etched YIG and conformal gold electrodes. Due to the combination of bi-layer liftoff and angled metal deposition, the gold electrodes are larger than designed and show a tapered thickness.

	\section{Experimental Results}
	
	\begin{figure}[!t]
		\vspace*{-0.2in}
		\centering
		\includegraphics[width=3.2in]{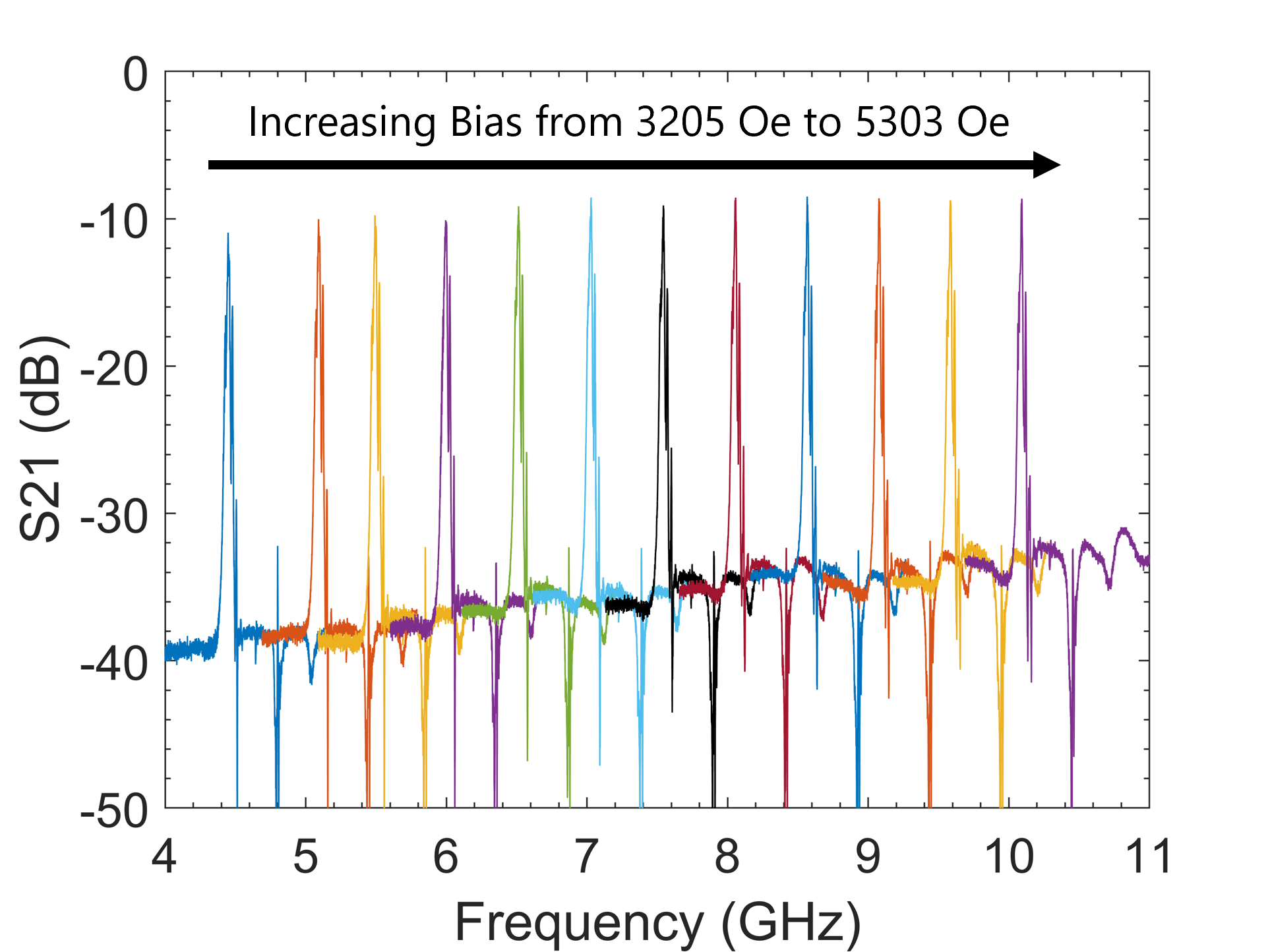}
		\caption{Measured 4-pole MSW bandpass filter frequency response at different out-of-plane magnetic biases from $\SI{3205}{Oe}$ to $\SI{5303}{Oe}$.}
		\label{fig_filter_frequency_response_tuning}
		\vspace*{-0.2in}
	\end{figure}

	\begin{figure}[b]
		\vspace*{-0.2in}
		\centering
		\includegraphics[width=3.2in]{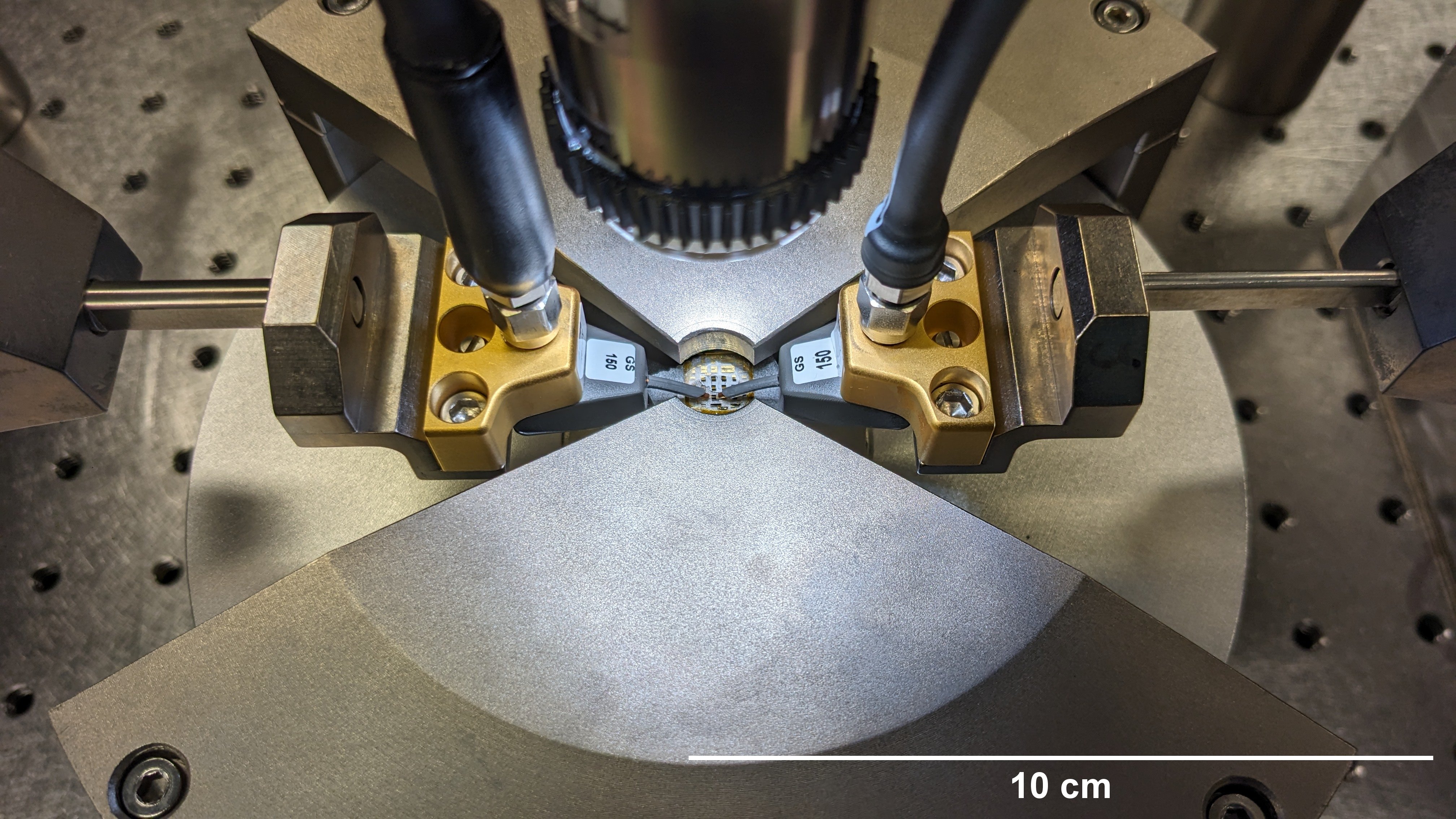}
		\caption{Experimental setup showing the fabricated filter chip resting on the pole of an electromagnet, two GS probes connected to one device under test, and an optical microscope used for probe landing and device alignment.}
		\label{measurement_setup}
	\end{figure}
	
	\begin{figure}[!t]
		\centering
		
		\vspace*{-0.2in}
		\subfloat[]{\includegraphics[width=3.2in]{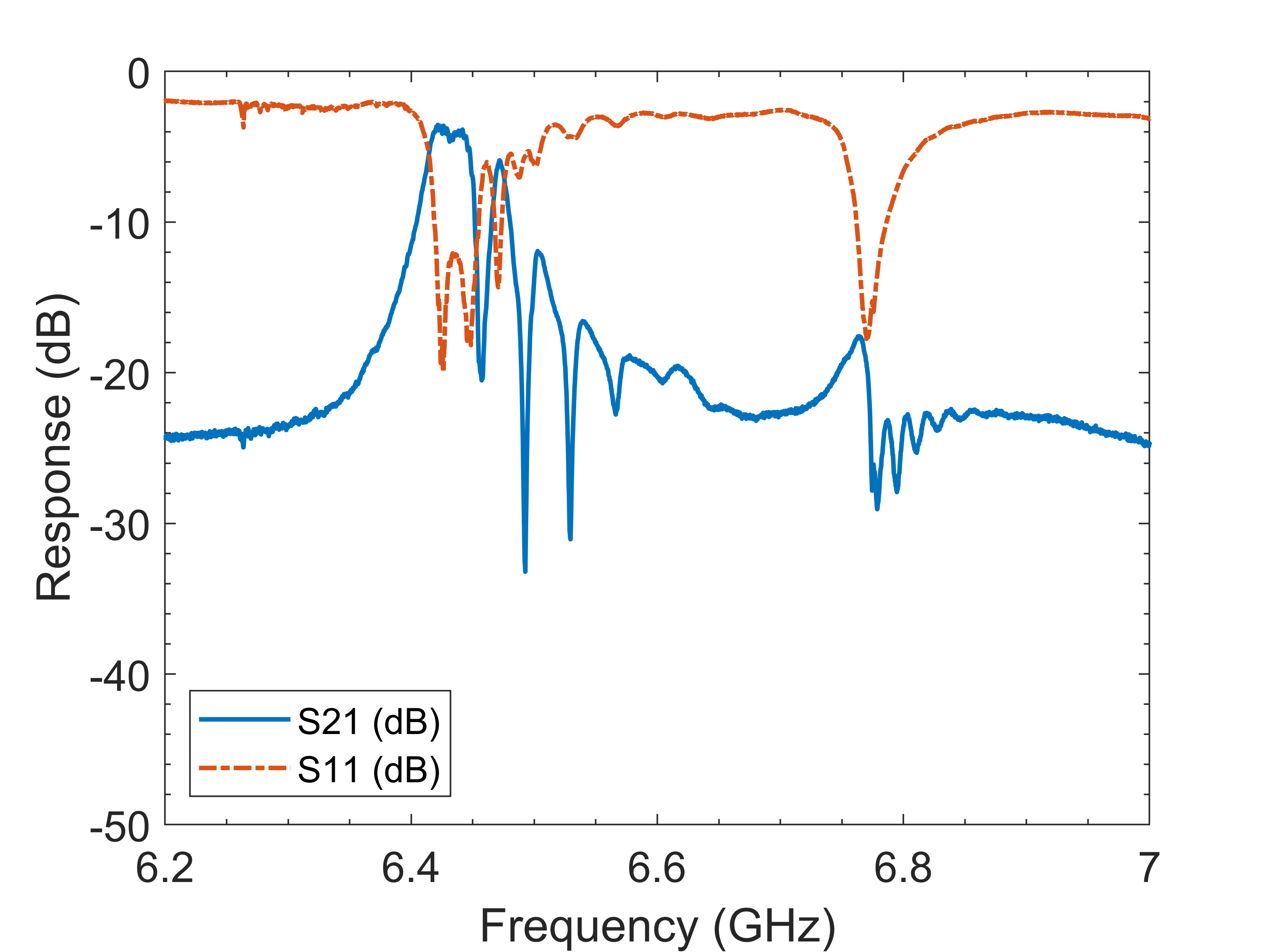}%
			\label{2pole_fig_filter_frequency_response}}
		
		\vspace*{-0.15in}
		\subfloat[]{\includegraphics[width=3.2in]{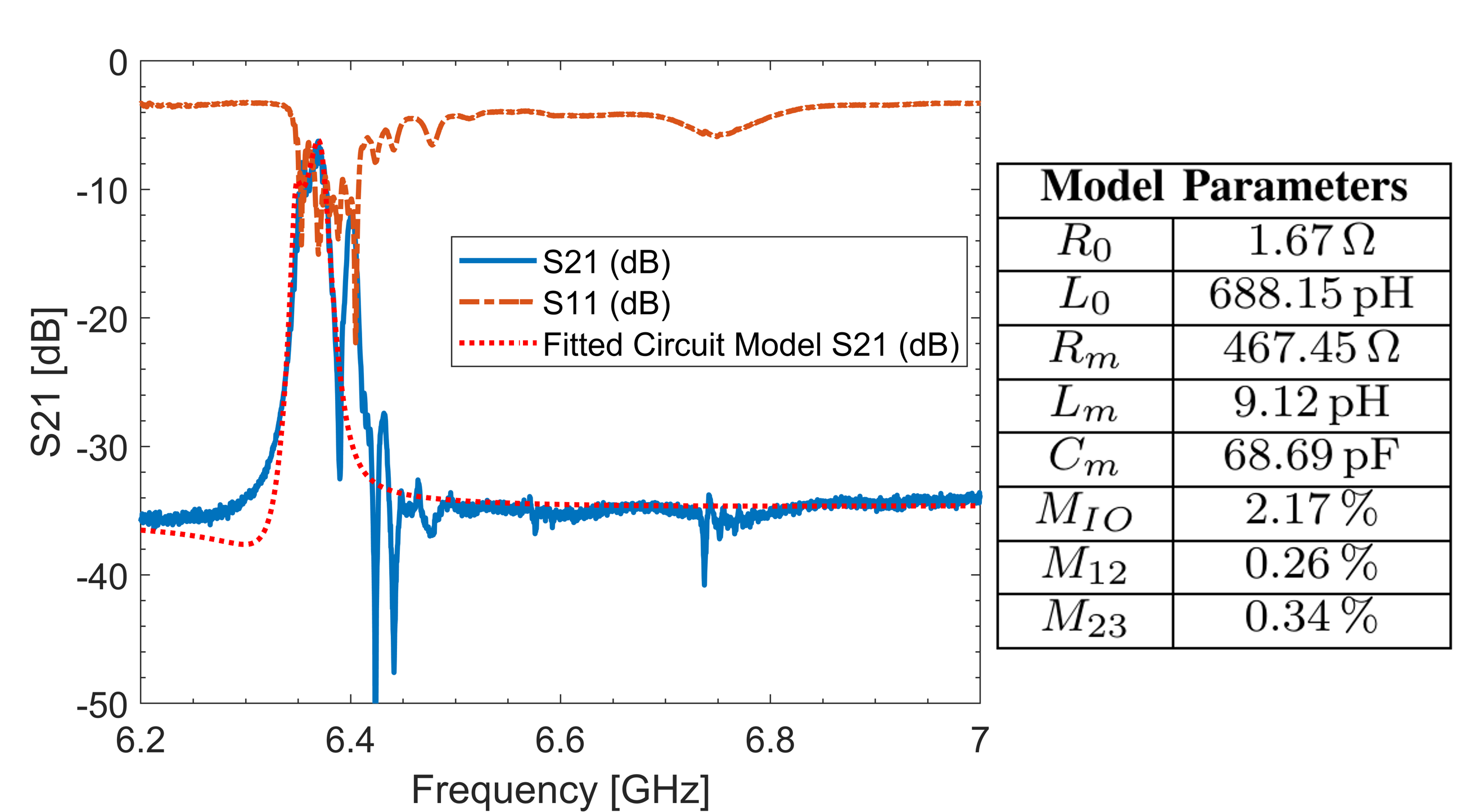}%
			\label{4pole_fig_filter_frequency_response}}
		
            \caption{Frequency responses for the \textbf{(a)} 2-pole and \textbf{(b)} 4-pole filters highlighted in Fig. $\ref{chip_pictures}$ near the passband at $\SI{3864}{Oe}$. \textbf{(b)} also shows a comparison of the fitted lumped element model in Fig. $\ref{lumped_circuit}$ with measured S21. The fitted circuit assumes all four resonators are identical and excludes the spurious pass bands caused by higher order MSFVW modes.}
            \vspace*{-0.2in}
		\label{frequency_response}
	\end{figure}

    \begin{figure}[!b]
        \vspace*{-0.2in}
		\centering
		\includegraphics[width=3.2in]{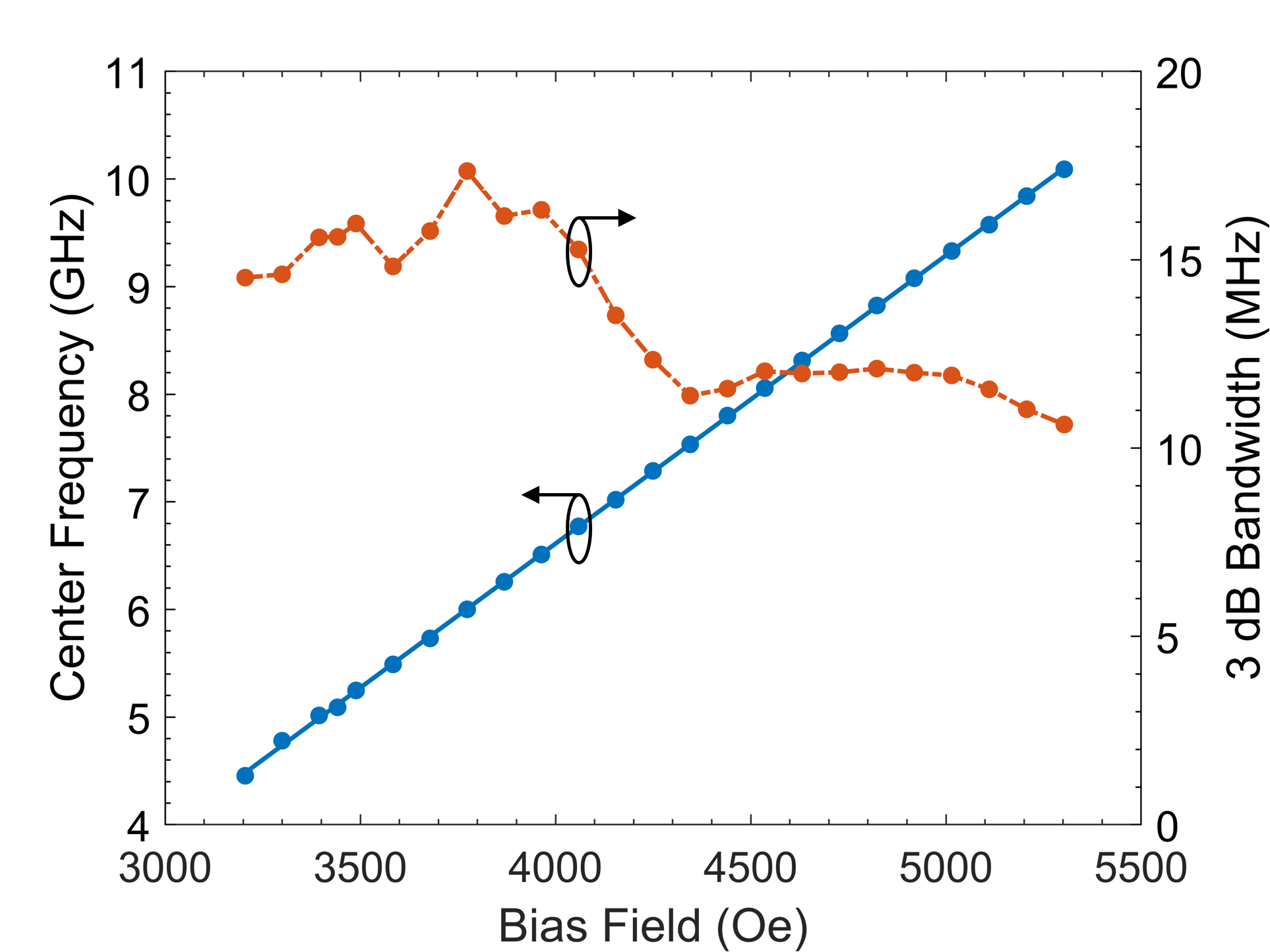}
		\caption{Measured $\SI{3}{\decibel}$ bandwidth and center frequency of the 4-pole MSFVW filter showing a tuning rate of $\SI{2.7}{\mega\hertz / Oe}$.}
		\label{fig_filter_tuning}
	\end{figure}
 
	Filter s-parameters are measured using an Agilent PNA-L N5230A network analyzer with a pair of ground-signal (GS) probes from $\SI{3205}{Oe}$ to $\SI{5303}{Oe}$ corresponding to a center frequency tuning over $\SI{5.6}{\giga\hertz}$ as shown in Fig. $\ref{fig_filter_frequency_response_tuning}$.\ignore{A stack of $\SI{1}{in}$ diameter, grade N52 Neodymium permanent magnets attached to a three axis stage provides the required out-of-plane magnetic field and a three axis hall effect sensor (THM1176-MF) is used to calibrate the applied field prior to each measurement.} A single-pole electromagnet powered by a constant current source provides the required out-of-plane magnetic bias (pictured in Fig. $\ref{measurement_setup}$) and a single-axis Gauss meter is used to map the source current to the applied field. Prior to each measurement, the device under test is aligned to the center of the electromagnet's pole to ensure field uniformity. Fig. $\ref{frequency_response}$ shows the frequency response near the passband for a 2-pole and 4-pole filter at $\SI{3864}{Oe}$. Outside of the spin wave manifold, no magnetostatic waves may propagate so the filter behaves as two coupled inductors. Consequently, the out-of-band rejection is governed by the inductive coupling strength between the input and output electrodes. For the 2-pole filter with an electrode spacing of $\SI{67}{\micro\meter}$, the rejection is  $\SI{-25}{\decibel}$ while for the 4-pole filer with a spacing of $\SI{229}{\micro\meter}$, the rejection is $\SI{-35}{\decibel}$. At $\SI{3864}{Oe}$, the 2-pole filter shows an insertion loss (IL) of $\SI{-3.55}{\decibel}$ and a $\SI{3}{\decibel}$ bandwidth of $\SI{57.0}{\mega\hertz}$ while the 4-pole filter has an IL of $\SI{6.94}{\decibel}$ and a $\SI{3}{\decibel}$ bandwidth of $\SI{17.0}{\mega\hertz}$. From Fig. $\ref{2pole_fig_filter_frequency_response}$ and $\ref{4pole_fig_filter_frequency_response}$, higher-order magnetostatic spurious modes are visible to the right of the passband with a frequency separation of $\SI{29}{\mega\hertz}-\SI{40}{\mega\hertz}$. As described in \cite{MSFVW:Ishak}, the current distribution along the transducer length can excite either even or odd ordered modes. Based on the resonator dimensions and electrically short transducer, the frequency spacing of these spurs agree well with odd ordered length modes. Fig. $\ref{fig_filter_tuning}$ shows the linear center frequency tuning at a rate of $\SI{2.7}{\mega\hertz/Oe}$ and the $\SI{3}{\decibel}$-bandwidth over the applied bias for the 4-pole filter. With a well-calibrated bias field, the extrapolated center frequency tuning line should intersect $-\omega_m$ at $\SI{0}{Oe}$. However, the Gauss meter used for the field calibration is thicker than the fabricated chip, so the reported bias is underestimated by approximately $\SI{219}{Oe}$. The filter's tuning was also measured using a neodymium permanent magnet mounted on a 3-axis stage to precisely calibrate the field accounting for any thickness difference between the chip and sensor. In this setup, the extrapolated $\SI{0}{Oe}$ intersection is at $\omega_m= \mu_0\gamma_m \cdot \SI{1751}{Oe}$ which agrees well the saturation magnetization of LPE YIG.

    \begin{figure}[!b]
		\centering
		
		\subfloat[]{\includegraphics[width=3.2in]{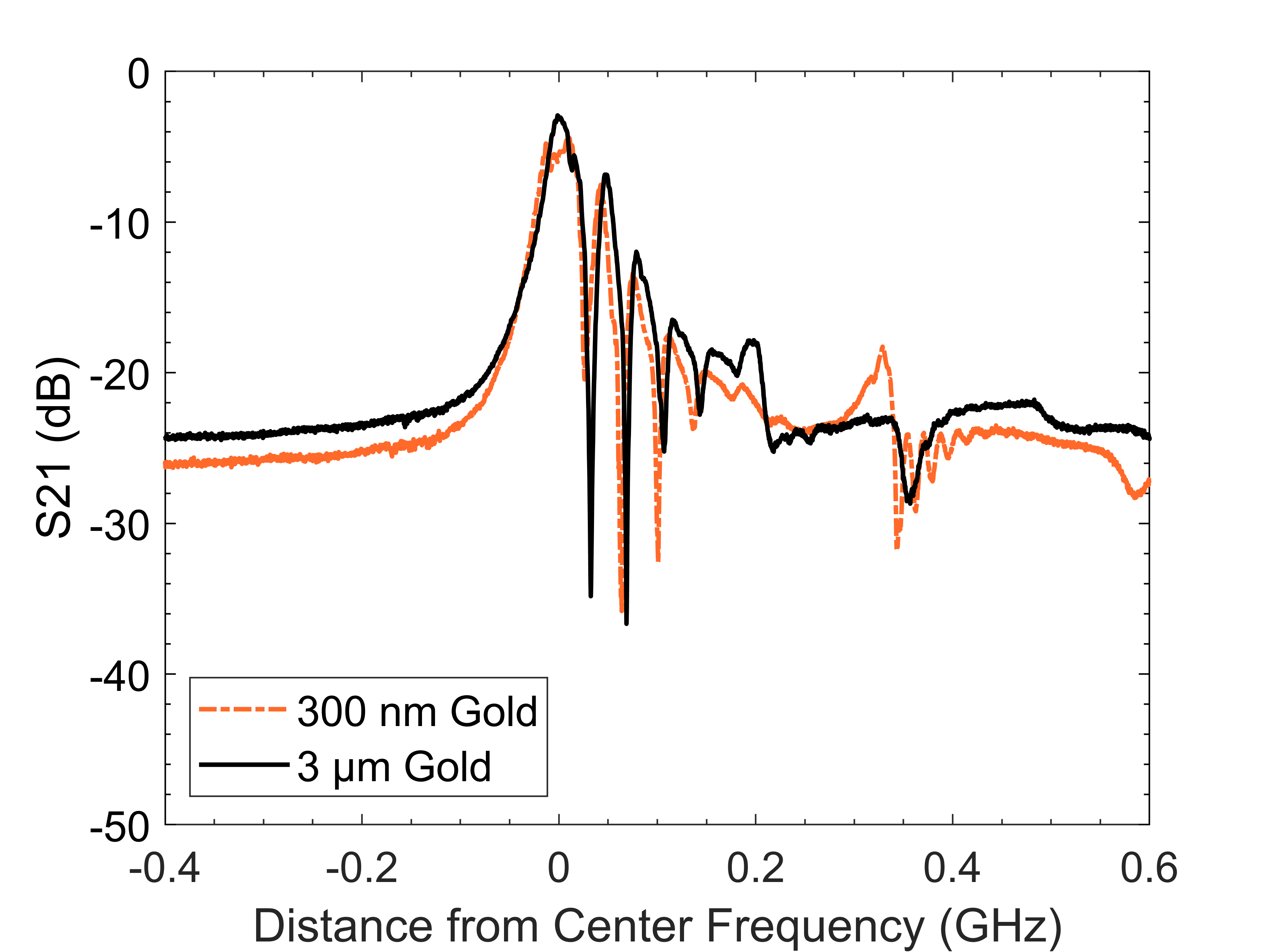}%
			\label{ElectrodeThickness_S21}}

        \vspace*{-0.15in}
		\subfloat[]{\includegraphics[width=3.2in]{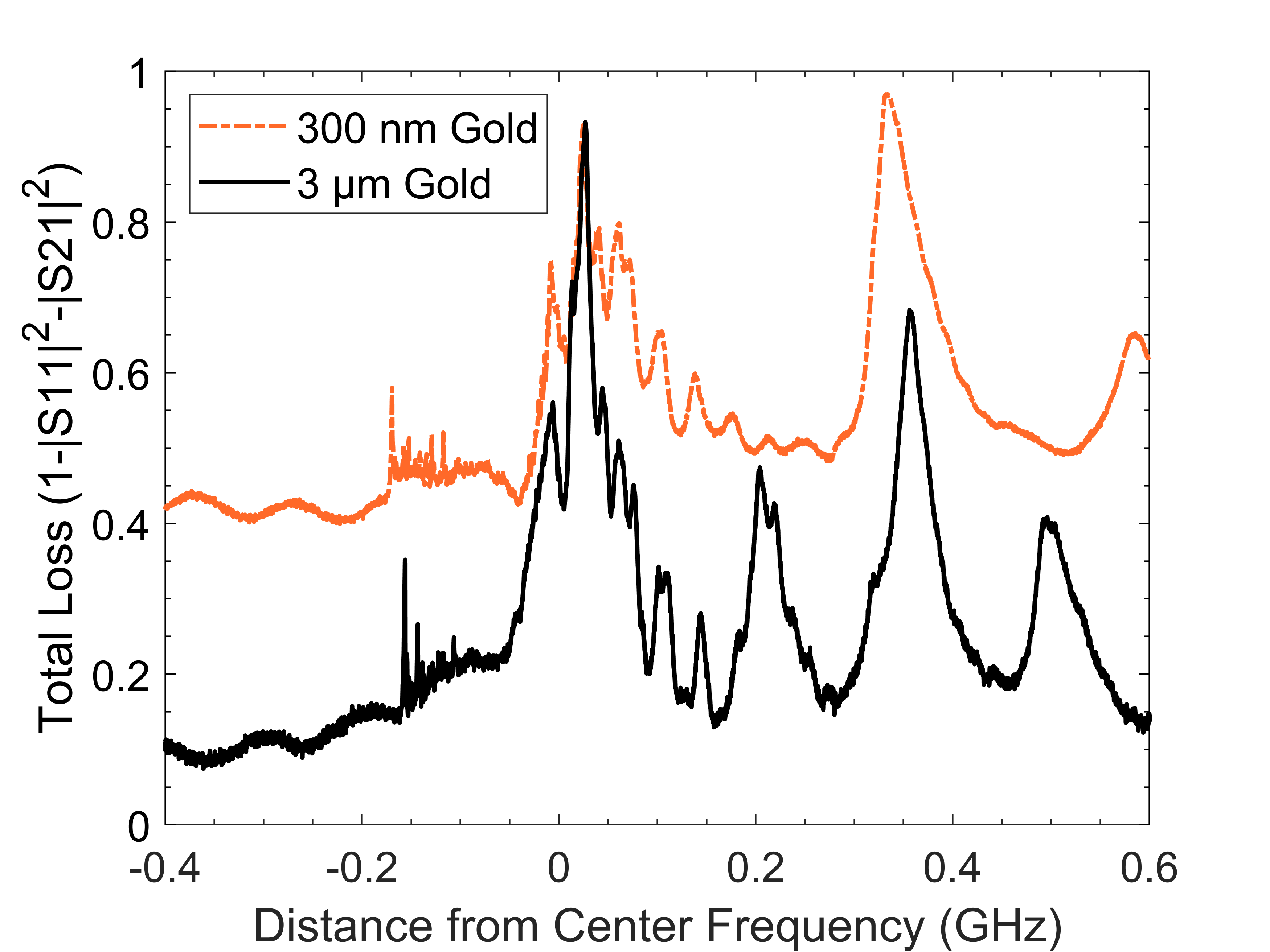}%
			\label{ElectrodeThickness_Loss}}
		
		\caption{\textbf{(a)} Measured S21 and \textbf{(b)} total loss for 2-pole filters with $\SI{300}{\nano\meter}$ and $\SI{3}{\micro\meter}$ thick gold transducers biased at $\SI{3652}{Oe}$ and $\SI{3660}{Oe}$ respectively. Frequency is plotted relative to the center frequency to account for the slight difference in bias strength.}
		
		\label{ElectrodeThickness}
	\end{figure}

    Considering the total loss ($1-\left\lvert S_{11}\right\rvert^2-\left\lvert S_{21}\right\rvert^2$) for the 2-pole filter biased at $\SI{3652}{Oe}$ and measured far away from all magnetostatic resonances, an average of $43\%$ of the input power is dissipated in the thin $\SI{300}{\nano\meter}$ gold transducers, radiated, or absorbed by the YIG on GGG substrate. Based on finite element simulations, the loss is primarily attributed to the resistance of the gold transducers. A second sample was fabricated with $\SI{3}{\micro\meter}$ electroplated gold to reduce resistive losses. Fig. $\ref{ElectrodeThickness}$ compares the measured insertion loss and total loss for the same 2-pole filter with different gold thicknesses. As expected, the mean out-of-band loss shows significant improvement from $43\%$ to only $12\%$. The average loss within the 3dB bandwidth exhibits a slight improvement dependent on bias, ranging from $0.5\%$ to $14.5\%$. The insertion loss improvement reflects the change in mean in-band loss with a maximum improvement from $\SI{4.43}{\decibel}$ to $\SI{2.92}{\decibel}$ around $\SI{3660}{Oe}$.

    \begin{figure}[!t]
		\vspace{-0.15in}
		\centering
		\includegraphics[width=3.2in]{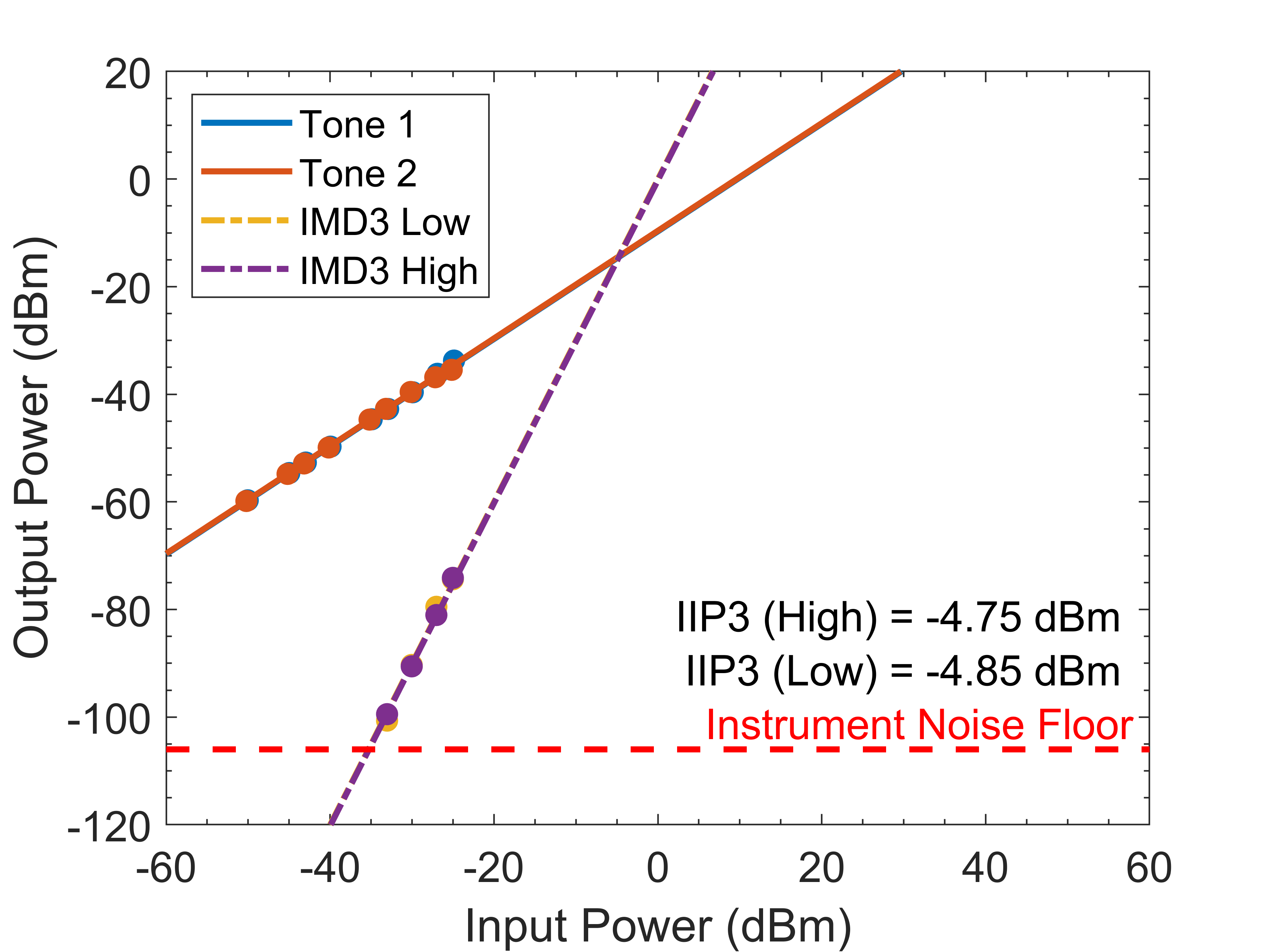}
		\caption{Two-tone IIP3 measurement in the passband of a 4-pole filter at $\SI{3652}{Oe}$ bias}
		\label{passband_IIP3}
		\vspace{-0.1in}
	\end{figure}
	
    \begin{table}[!t]
		\caption{Upper Input tone frequencies for IIP3 measurements}
		\centering
		\renewcommand{\arraystretch}{1.5}
		\begin{tabular}{|c|c|c|c|}\hline
			Bias Field  & Stopband Low & Passband & Stopband High \\ \hline
			3652 Oe & $\SI{5.465}{\giga\hertz}$ & $\SI{5.799}{\giga\hertz}$ & $\SI{6.055}{\giga\hertz}$ \\\hline
		\end{tabular}
		\label{IIP3_summary_frequency_table}
	\end{table}
 
    \begin{table}[!t]
		\caption{Summary of 4-pole filter IIP3}
		\centering
		\renewcommand{\arraystretch}{1.5}
		\begin{tabular}{|c|c|c|c|}\hline
			Bias Field  & Stopband Low & Passband & Stopband High \\ \hline
			3652 Oe & $\geq \SI{37.95}{dBm}$ & $\SI{-4.85}{dBm}$ & $\SI{25.84}{dBm}$ \\\hline
		\end{tabular}
		\label{IIP3_summary_table}
	\end{table}

    The linearity of the MSW bandpass filter is evaluated by measuring the input referred $3^{rd}$ order intercept point (IIP3) in the passband as well as the stopband both below and above the passband at a bias of $\SI{3652}{Oe}$. The nonlinearity measurements are performed using two Keysight E8257D signal generators with a frequency separation of $\Delta f = \SI{15}{\mega\hertz}$. The higher of the two tone frequencies for each region are listed in Table $\ref{IIP3_summary_frequency_table}$. A wideband power divider combines the two tones while an Agilent PXA spectrum analyzer measures the resultant spectrum. A two-stage calibration is performed to remove all cable and system losses at every tone frequency and input power level. Far away from the passband, the filter is expected to be linear and no intermodulation products were observed. Based on the maximum output power of the signal generators and noise floor of the spectrum analyzer, the lower stopband IIP3 is estimated to be greater than $\SI{37.95}{dBm}$ at $\SI{3652}{Oe}$. The passband shows the greatest nonlinearity with an IIP3 of $\SI{-4.75}{dBm}$ at $\SI{3652}{Oe}$ as shown in Fig. $\ref{passband_IIP3}$. The measured IIP3 in each frequency region is summarized in Table $\ref{IIP3_summary_table}$.

	\section{Conclusion}

	\begin{table} [t]
		\caption{Performance Comparison with Other Tunable Bandpass Filters}
		\centering
		\renewcommand{\arraystretch}{1.5}
		\begin{tabular}{|p{2.2cm}|p{1.2cm}|p{1.1cm}|p{1.1cm}|p{1.1cm}|}\hline
			Reference  & Frequency Tuning (GHz)  & Insertion Loss (dB) & Bandwidth (MHz) & Rejection (dB) \\\hline
			This work (2-pole) & 4.5-10.1 & $<6$ & 29-39 & $>25$ \\\hline
			This work (4-pole) & 4.5-10.1 & $<11$ & 11-17 & $>35$ \\\hline
			YIG Sphere \cite{micro_lambda_wireless_inc_mlfd_nodate} &  2-8 & $<5$  &  $20$  & $>50$  \\\hline
			Magnetostatic Surface Wave \cite{du_frequency_2023}  & 3.4-11.1 & $<5.1$ & 18-25 & $>25$ \\\hline
            RF MEMS Tunable Filter \cite{entesari_differential_2005} & 6.5-10 & $<5.6$ & 306-539  & $>50$ \\\hline
            Evanescent-Mode Cavity \cite{joshi_multi_2010, joshi_highly_2007}  & 3-6.2 & $<4$ & 15-25 & $>50$ \\\hline
		\end{tabular}
		\label{Performance_Comparison}
	\end{table}

	In this paper, we have demonstrated a novel edge-coupled highly-tunable magnetostatic bandpass filter using state-of-the-art micromachining fabrication techniques. The designed 2-pole and 4-pole filters have been tuned over an octave from $\SI{4.5}{\giga\hertz}$ to $\SI{10.1}{\giga\hertz}$ showing a consistent passband shape with performance comparable to other state-of-the-art frequency tunable bandpass filters as summarized in Table $\ref{Performance_Comparison}$. We have also characterized the linearity of the filter in three distinct frequency regions. Our micromachining process enables precise control over the YIG mesa shape and spacings to synthesize miniaturized MSW channel-select filters analogous to electromagnetic cavity filter design.
	
	\section*{Data Availability}
	The code and data used to produce the plots within this work will be released on the repository Zenodo upon publication. 
	
	\section*{Acknowledgments}
    Chip fabrication was performed at the Birck Nanotechnology Center at Purdue. Resonator and filter measurements and characterization were performed at Seng-Liang Wang Hall at Purdue. The Purdue authors would like to thank Dave Lubelski for assistance with the glancing angle metal deposition and Yiyang Feng for discussions on the fabrication recipes.

	\balance
	
	\bibliographystyle{IEEEtran}
	\bibliography{MSFVW_bib}
	
	\vfill
	
\end{document}